\documentclass[preprint,superscriptaddress,floatfix]{revtex4-2}
\usepackage[utf8]{inputenc}
\usepackage[T1]{fontenc}
\usepackage{lmodern} 
\usepackage{graphicx} 
\usepackage{graphics}
\usepackage{epsfig}
\usepackage{epstopdf}
\usepackage{dcolumn}
\usepackage{multirow}
\usepackage{amsmath,amssymb} 
\usepackage{enumitem}
\usepackage{fullpage}
\usepackage{tabularx,longtable,booktabs,diagbox}

\usepackage[english]{babel}
\usepackage[colorlinks=true,linkcolor=blue,citecolor=magenta,urlcolor=blue]{hyperref}

\begin{document}

\title{Energy extraction from a Kerr-Bardeen-Letelier black hole by Comisso-Asenjo Mechanism}

\author{Bijendra Kumar Vishvakarma}
\email{bkv1043@gmail.com}
\affiliation{Department of Physics, Banaras Hindu University, Varanasi - 221005, Uttar Pradesh, India.}

\author{Dharm Veer Singh}
\email{veerdsingh@gmail.com}
\affiliation{Department of Physics, Institute of Applied Sciences and Humanities, GLA University, Mathura - 281406, Uttar Pradesh, India. 
{\footnote{Visiting Associate: IUCAA, Pune, Maharashtra  and School of Physics, Damghan University, Iran. }}}
\author{Sanjay Siwach}
\email{sksiwach@hotmail.com}
\affiliation{Department of Physics, Banaras Hindu University, Varanasi - 221005, Uttar Pradesh, India.}
\begin{abstract}
We explore the energy extraction from a Kerr-like regular black hole  coupled with a cloud of strings (CoS) via the Comisso-Asenjo mechanism.  We study the effects of the black hole spin ($a$), black hole charge ($g$) and CoS parameter ($b$) on the ergo-region of the black hole and the magnetic reconnection process, which is governed by the Comisso-Asenjo mechanism. We also analyze the phase space involving negative energy density of the relativistic magnetized plasma at infinity. The rate of energy extraction in the magnetic reconnection process is also computed and compared with that of the Blandford-Znajek mechanism. 
\end{abstract}

\maketitle

\newpage

\section{\label{sec:level1}Introduction}

Black holes provide an interesting laboratory for testing extreme gravity and the ideas of quantum gravity and their direct detection have opened up a new window for astrophysical observations \cite{eht1}. The gravitational wave produced by black hole mergers is established beyond doubt by ground based large interferometer based observations \cite{lv1234}. The  emission of high energy jets of particles and gamma-ray bursts has also opened up new frontiers for research in black hole physics \cite{1,2,3}. The loss of energy by a black hole by various mechanisms is a subject of investigation and the Penrose process of energy extraction is well known in the literature \cite{Wagh1985,Pradhan}. Based on Penrose’s work, several researchers have proposed various means of extracting energy from black holes, such as the collisional Penrose process \cite{rp6}, the Blandford–Znajek process  in the magnetic field \cite{Kshitij,rp8} and the magnetohydrodynamic process \cite{rp9, rp10}. Among these, the Blandford–Znajek process has gained wide acceptance for explaining the energetic jets and gamma-ray bursts observed in active galactic nuclei. In this process, the interaction of the charged particle with the surrounding magnetic field generates a current, thereby twisting the magnetic field lines into a helical structure, resulting in energy extraction from the black hole that is channeled into relativistic jets  \cite{rp11,rp12,rp13,rp14,rp15,rp16,rp17}.

Koide and Arai \cite{mr1} were the first to explore magnetic reconnection as a possible energy extraction mechanism from a rotating black hole. This  process facilitates the conversion of magnetic field  energy into kinetic energy carried by surrounding plasma \cite{mr2,mr3,mr4,mr5}. The energy extraction via the Comisso-Asenjo mechanism was first studied in the context of a Kerr black hole \cite{Comisso21}, and it has been extended to many black holes in recent years \cite{ca1,ca2,ca3,ca4,ca5}.
Several works have been done related to the energy extraction from spinning black holes via magnetic reconnection \cite{khodadi2022magnetic,carleo2022energy,chen2024energy,zeng2025energy}. This mechanism is similar to the Penrose process, and magnetic reconnection generates accelerated plasma flux in opposite directions. The outflow is driven against the black hole spin, thus enabling the energy extraction from the black hole. The astrophysical observations seem to lend support to this mechanism as a potential source of energy extraction  from a black hole \cite{tchekhovskoy2008simulations,tchekhovskoy2011efficient,neilsen2015xray,abuter2021constraining}. The work is also generalized for non-Kerr black holes for energy extraction via magnetic reconnection process \cite{wang2022extracting,li2023energyhairy,li2023energyregular,khodadi2023harvesting,shaymatov2023kerrnewman,zhang2024energy}.

In this work, we aim to investigate the energy extraction from the Kerr-like regular black hole coupled with the Cloud of Strings (CoS) via Comisso-Asenjo mechanism. The regular black hole solutions are obtained by coupling gravity with nonlinear electrodynamics (NED).  The first regular black hole model, proposed by Bardeen \cite{Bardeen:1968}, was based on the Gliner   and Shakarov proposal \cite{gil, AD} and later identified as a solution of gravity coupled with NED by  Ayno Beato and Garcia  \cite{ABG99}. Letelier obtained a black hole solution in the cloud of strings in the universe \cite{letelier1979clouds}. There are many black hole solutions based on the Bardeen proposal \cite{Singh:2022xgi} 
and the regular black hole solutions coupled with CoS \cite{Sudhanshu:2024wqb,dvs22,santos2025,Rodrigues:2022zph,kumar2024ayon}.    

The paper is organized as follows. In section 2, we consider a Kerr-like Bardeen black hole in a CoS and calculate the radius of photon orbits using geodesic equations. In section 3, we review the details of the process of magnetic reconnection. In section 4, we consider the fluid approximation of energy momentum tensor required for magnetohydrodynamic equations, and the energy density at infinity in the Zero Angular Momentum (ZAMO) frame.  In section 5, the efficiency of energy extraction process is examined and results are compared with that of the Blandford-Znajek mechanism. 
 %%%%%%%%%%%%%%%%%%%%%%%%%%%%%%%%
\section{\label{sec:level2}SPINNING BARDEEN BLACK HOLE IN CLOUD OF STRINGS} 
Let us consider the action of Einstein's gravity coupled with an NED and CoS source \cite{Vishvakarma:2023tnl},  
\begin{equation}
S=\int d^4x \sqrt{-g}\left[R+{\cal L}_{NED}+{\cal L}_{CoS}\right],
\label{action}
\end{equation}
where $R$ is the Ricci scalar, ${\cal L}_{NED}$ is the  Lagrangian densities of the nonlinear source and ${\cal L}_{CoS}$ is the Lagrangian densities of the CoS source. 

Lagrangian density ${\cal L}_{NED}$ expressed in terms of Maxwell scalar $F=\frac{1}{4}F_{\mu\nu}F^{\mu\nu}$ as
\begin{equation*}
    {\cal L}_{NED}=\frac{3}{2sg^2}\left [\frac{\sqrt{2 g^2 F}}{1+2 g^2 F}\right]^\frac{5}{2}
\end{equation*}

where $s=g/2M$. The equations of motion are obtained by varying the action (\ref{action}) with respect to the metric tensor, $g_{\mu\nu}$ and electromagnetic potential, $A_{\mu}$, and can be written as,  
\begin{eqnarray}
&& R_{\mu\nu}-\frac{1}{2}g_{\mu\nu}R=T_{\mu\nu}^{NED}+T_{\mu\nu}^{CS},\\
&& \nabla_{\mu}\left(\frac{\partial {\cal L}}{\partial F}F^{\mu\nu}\right)=0
\label{eom}
\end{eqnarray}

The energy-momentum tensor (EMT)  of the nonlinear source is given by,  
\begin{eqnarray}
&&  T_{\mu\nu}^{NED}=2\left[\frac{\partial {{L(F)}}}{\partial F}F_{\mu \gamma}F_{\nu}^{\gamma}- g_{\mu\nu}{{L(F)}}\right]
%=\frac{8M g^2}%{(r^2+g^2)^{5/2}},
\label{emt}
\end{eqnarray}
 
The CoS term in the action  (\ref{action}) is given by the Nambu-Goto action, and the stress tensor is written as, \cite{letelier1979clouds} 
\begin{equation}
T_{\mu \nu}^{CoS}  = \frac{\rho \Sigma^{\mu \rho} \Sigma_{\rho}^{\phantom{\rho} \nu}}{\sqrt{-\gamma}}
%=\frac{b}{r^2},
\end{equation}
where $\rho$ is a density, $\Sigma^{\mu \nu} $ is a bivector and $-\gamma$ is determinant of world-sheet metric $\gamma_{ab}$ \cite{Rodrigues:2022embedding}.

The above equations admit a black hole solution, known as the Bardeen-Letelier black hole \cite{Rodrigues:2022zph}, and the line element is given by 
\begin{equation}
ds^2=-\left(1-\frac{2Mr^2}{(r^2+g^2)^{3/2}}-b\right) dt^2+\frac{1}{\left(1-\frac{2Mr^2}{(r^2+g^2)^{3/2}}-b\right)}+ r^2 d\theta^2+r^2\sin^2\theta d\phi^2
\label{solution}
\end{equation}
 
 The solution is characterized by mass ($M$), magnetic monopole charge ($g$), and a CoS parameter ($b$). The rotating counterpart of Bardeen-Letelier Black hole (\ref{solution}) is obtained by using the Newman-Janis procedure \cite{E24} and the line element is given by,  

\begin{eqnarray}
 ds^{2}=-\left(1-\frac{b\ r^2+\frac{2\ M\  r^4}{(r^2+g^2)^{3/2}}}{\Sigma}\right)dt^{2}-\frac{2\ a\sin^{2}\theta}{\Sigma}\left(r^{2}+a^{2}-\Delta\right) dt \ d\phi+\frac{\Sigma}{\Delta}dr^{2}+\nonumber\\
\Sigma \,d\theta^{2}  +\frac{\sin^{2}\theta}{\Sigma}\Bigg[(r^{2}+a^{2})^{2}-\Delta ~ a^{2}\sin^{2}\theta\Bigg]d\phi^{2},
\label{bhs}
\end{eqnarray}
where
\begin{eqnarray}
 \Delta=(1-b) r^{2}+a^{2}-\frac{2Mr^{4}}{(r^2+g^2)^{\frac{3}{2}}} \qquad\text{and}\qquad\Sigma = r^{2}+a^{2} \cos^{2}\theta
 \label{mat1}
\end{eqnarray}

The spin parameter, ($a$), is related with the angular momentum, ($J$) via the relation $a=J/M$. The Black hole solution (\ref{bhs}) with (\ref{mat1}) reduces to the Kerr-Letelier Black hole in the absence of magnetic monopole charge, the Kerr-Bardeen black hole in the absence of CoS parameter.

\begin{figure*}[ht]
\begin{tabular}{c c}
\includegraphics[width=0.45\textwidth]{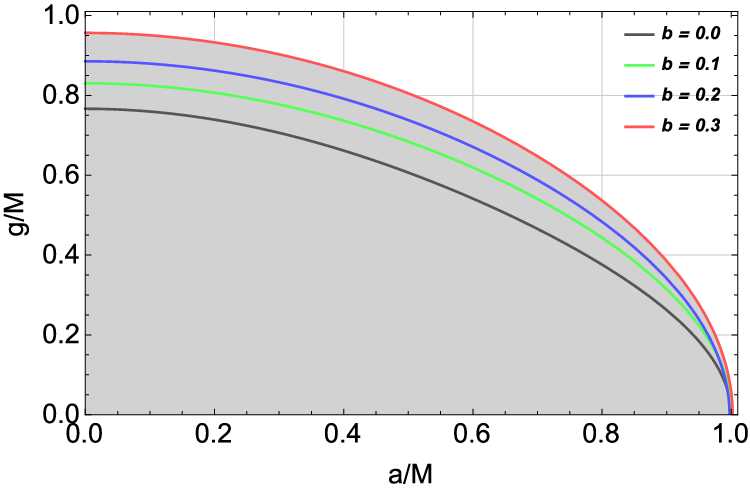}
\includegraphics[width=0.5\textwidth]{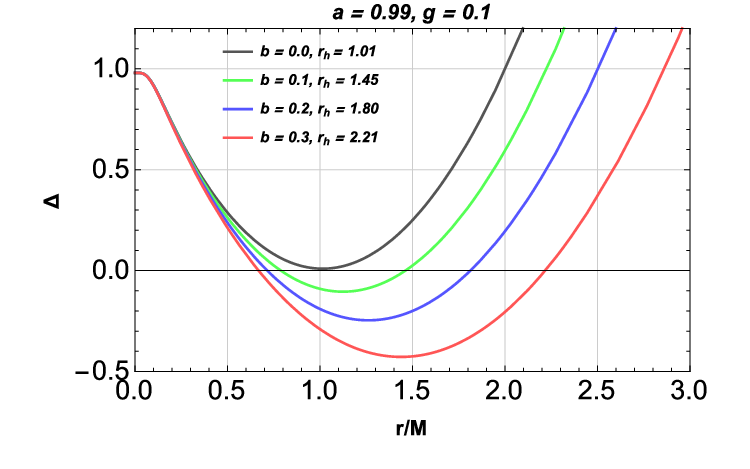}
\end{tabular}
\caption{Plot of parameter space ($g/M$ vs $a/M$) of Kerr-Bardeen-Letelier Black hole  for different values of CoS parameter ($b=0,~0.1,~0.2,~\& ~0.3$) (left) and plot of $\Delta$ vs $r/M$ (horizon of black hole) for different values of CoS parameter with fixed values of $a$ and $g$ (right). }
\label{Fig:1}
\end{figure*}
 
\begin{figure*}[ht]
\begin{tabular}{c c}
\includegraphics[width=.45\textwidth]{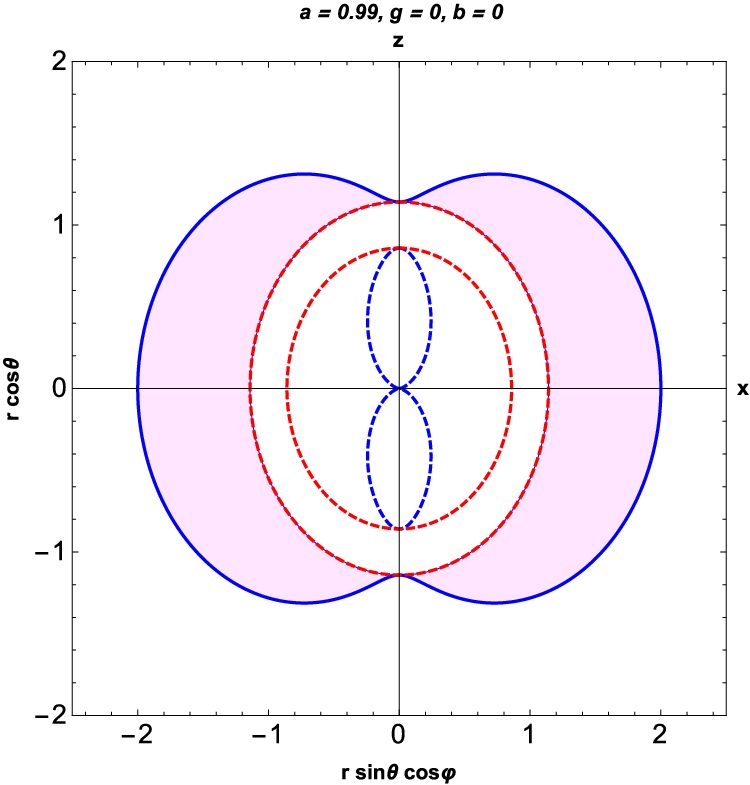}
\includegraphics[width=.45\textwidth]{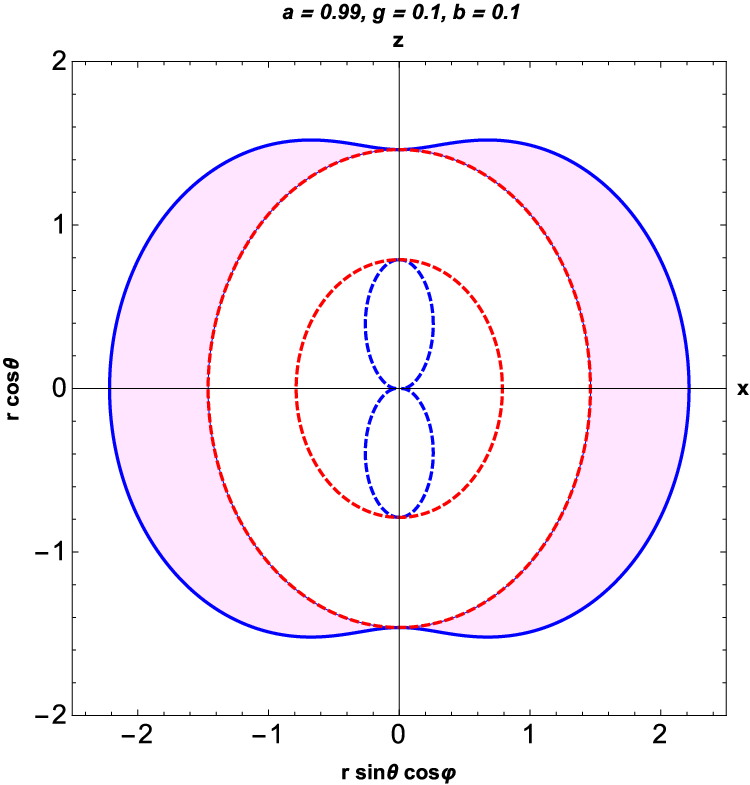}\\
\includegraphics[width=.45\textwidth]{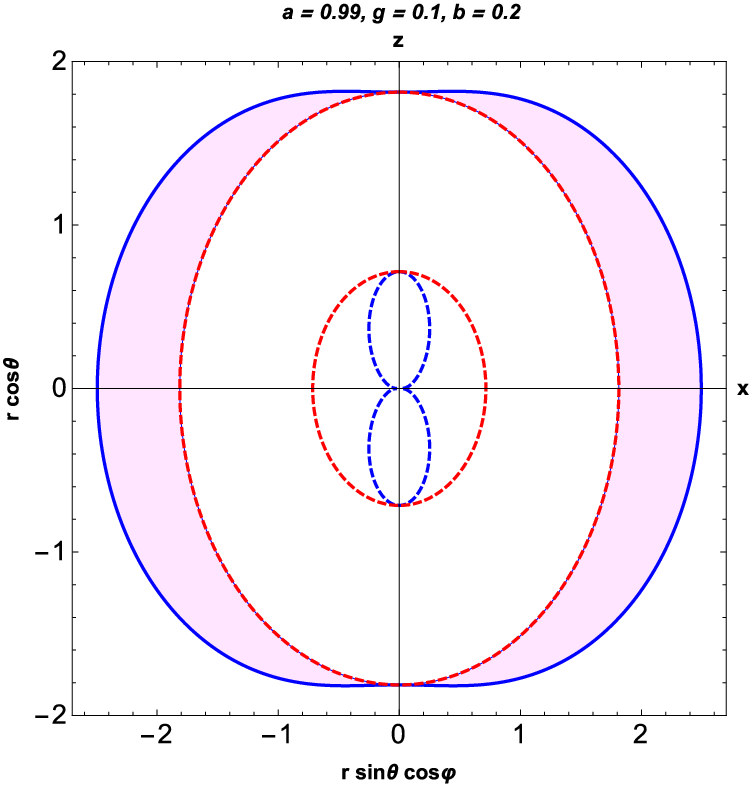}
\includegraphics[width=.45\textwidth]{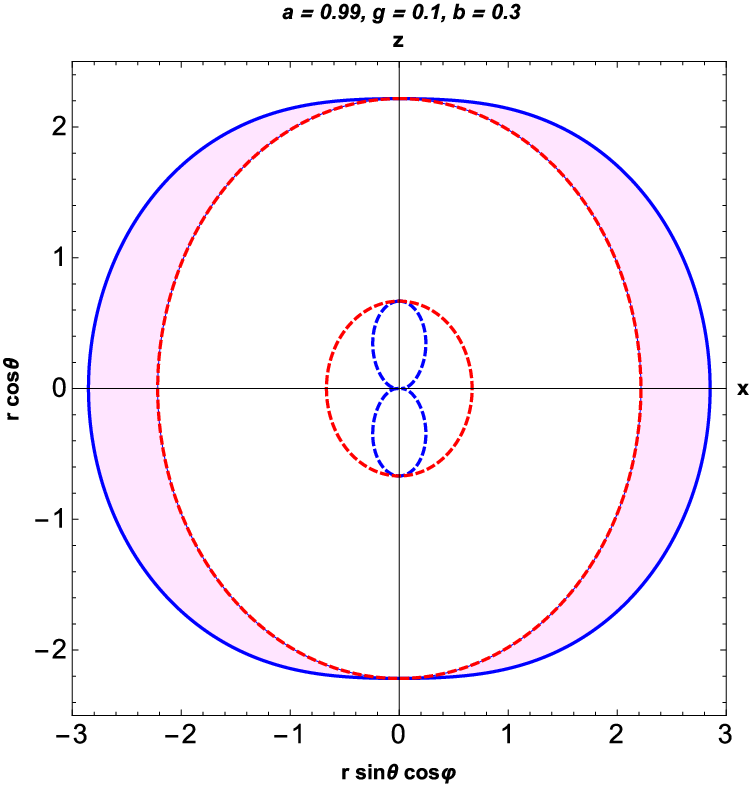}
\end{tabular}
\caption{Plot of ergo-regions of Kerr-Bardeen-Letelier black hole for different CoS parameter $b=0,\,0.1,\,0.2,\&~0.3$ with  fixed values of $a=0.99,g=0.1$. }
\label{Fig:2}
\end{figure*}

The parameter space of Kerr-Letelier-Bardeen black hole  is plotted in the Fig. \ref{Fig:1} for different values of CoS parameter, $b$. The static limit surface and ergo-region with different values $a,g,b$ are already obtained in our paper  \cite{Vishvakarma:2023tnl}. Here, we plot the ergo-region of the Kerr-Bardeen-Letailer black hole in Fig. \ref{Fig:2} for different values of the CoS parameter $(b)$  and fixed values of $(a,g$). We see that as we increase the value of the CoS parameter, $(b)$ the resulting ergo-region also increases and coincides at the pole. 
%%%%%%%%%%%%%%%%%%%%%%%%%%%%%%%%%
\section{\label{sec:level3} ENERGY EXTRACTION VIA COMISSO-ASENJO MECHANISM}
In this section, we review the Comisso-Asenjo mechanism of energy extraction from a black hole and apply it to the Kerr-Bardeen-Letelier black hole. The central theme of the Comisso-Asenjo mechanism is the extraction of energy from ergo-region of the black hole by magnetic re-connection. In this process the energy of the magnetized plasma surrounding the black hole is taken away as heat energy of the the particles entering the ergo-region. It is convenient to define various quantities in the ZAMO frame. The metric of the ZAMO frame is given by
\begin{equation}
   ds^{2}=-d\hat{t}^2 +\sum_{i=1}^{3}(d\hat{x^{i}})^{2}=\eta_{\mu \nu}d{\hat{x}}^{\mu}d{\hat{x}}^{\nu}
\end{equation}
Here, the hat represents the ZAMO frame, and $d\hat{t}=\alpha dt$, $d{\hat{x^i}}=\sqrt{g_{ii}}dx^i-\alpha \beta^i dt$. The $\alpha$ and $\beta$ are given by,
\begin{equation}
   \alpha=\sqrt{\left(-g_{tt}+\frac{g_{\phi t}^2}{g_{\phi\phi}}\right)}\qquad \text{and} \qquad \beta^i= \frac{\sqrt{g_{\phi\phi}}\omega^{\phi}}{\alpha},
\end{equation}
where $\omega^{\phi}$ is the angular velocity. 

The EMT of the system includes the contribution from both the electromagnetic energy as well as energy of the plasma fluid and can be conveniently expressed as,  
\begin{equation}
    T^{\mu\nu}=pg^{\mu \nu}+wU^{\nu}U^{\mu}+F^{\mu}_{\delta}F^{\nu \delta}-\frac{1}{4}g^{\mu \nu}F^{\rho \delta}F_{\rho \delta}.
    \label{eqn:13}
\end{equation}
where, $p$ is proper plasma pressure, $w$ is an enthalpy density, $U^{\mu}$ is the  four velocity, and $F^{\nu\delta}$ is the electromagnetic field tensor, respectively. Using this EMT ( Eq. \ref{eqn:13}), we can get the energy density at infinity $ e^{\infty} = e^{\infty}_{\text{Emd}} + e^{\infty}_{\text{Hyd}}$, where
\begin{eqnarray}
   e^{\infty}_{\text{Hyd}} = \alpha \hat{e}_{\text{Hyd}} + \alpha \beta^{\phi} w \gamma^{2} \hat{v}^{\phi} \qquad\text{and}\qquad e^{\infty}_{\text{Emd}} = \alpha \hat{e}_{\text{Emd}} + \alpha \beta^{\phi} (\hat{B} \times \hat{E})_{\phi} \label{eqn:16c}
\end{eqnarray}
with $\hat{e}_{\text{Emd}}= (\hat{E}^2+\hat{B}^2)/2$ and $\hat{e}_{\text{Hyd}}=w \hat{\gamma}^2-p$,  are the energy densities of  electromagnetic and hydrodynamic in ZAMO frame and $\hat{\gamma} = \hat{U}^0 = \left(1 - \sum_{i=1}^{3} (d\hat{v}^i)^2 \right)^{-1/2}$ is the Lorentz factor.

We assume that the electromagnetic energy is converted into  the  particle kinetic energy of plasma in the Comisso-Asenjo reconnection process. So we can ignore the contribution of electromagnetic energy in the total energy. For energy extraction, the following conditions should be obeyed, 
\begin{eqnarray}
    \Delta e_{+}^{\infty}=e_{+}^{\infty}-\left(1-\frac{\Gamma}{\Gamma-1}\frac{p}{w}\right),\qquad    e_{-}^{\infty}<0,\qquad\Delta e_{+}^{\infty}>0,
     \label{eqn:15}
\end{eqnarray}
where, Eq. (\ref{eqn:15}) represents energy density at infinity by a relativistically hot plasma of polytropic index $\Gamma={4}/{3}$ and $cos\ \xi=\frac{1}{\sqrt{3}}$.  
\begin{eqnarray}
e_{+}^{\infty} \simeq \frac{\sqrt{\sigma}\ (2+b(1+g^2)^\frac{3}{2})}{(1+g^2)^\frac{3}{2}\sqrt{2+b+\frac{2}{(1+g^2)^\frac{3}{2}}}\ cos \xi} \quad \text{and} \quad 
e_{-}^{\infty} \simeq -\sqrt{\frac{1}{\sigma}} \frac{ (2+b(1+g^2)^\frac{3}{2}) \ cos\xi}{(1+g^2)^\frac{3}{2}\sqrt{2+b+\frac{2}{(1+g^2)^\frac{3}{2}}}}
 \end{eqnarray}
 
\begin{figure*}[ht]
\includegraphics[width=0.45\textwidth]{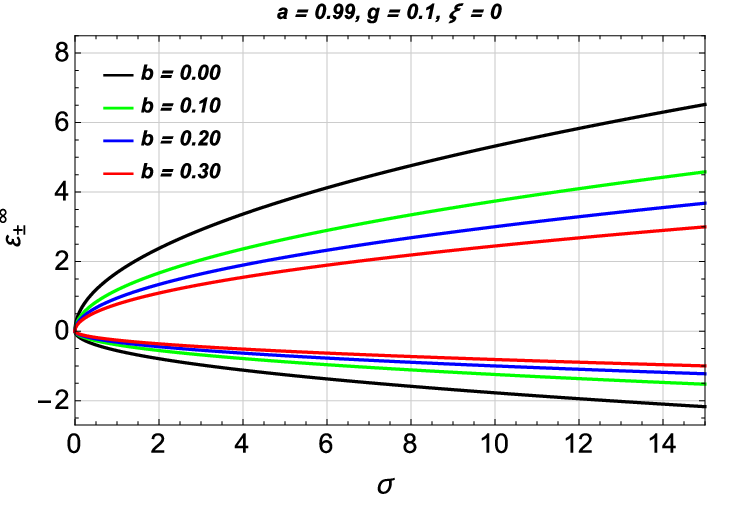}
\includegraphics[width=0.45\textwidth]{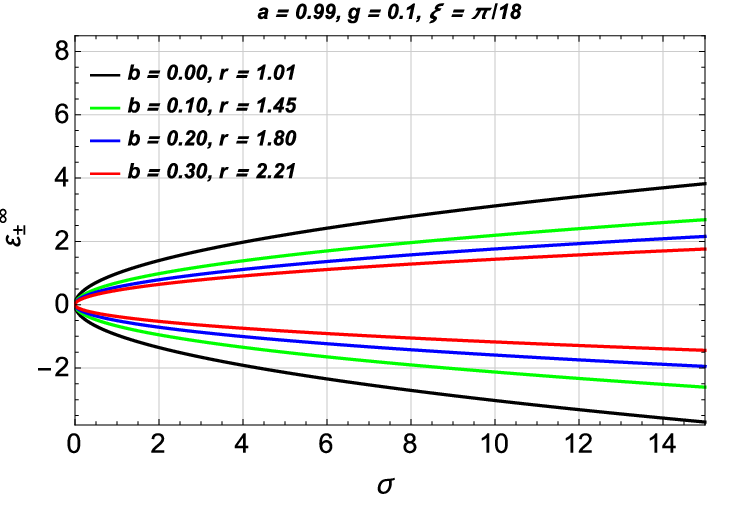}
\caption{ The plot of energy density at infinity per enthalpy vs plasma magnetization with fixed values of the CoS parameter ($b=0,0.1,0.2,\&\ 0.3$) respectively with fixed values of $a=0.99, g=0.1$, $\xi=0,\pi/18$ and X-point location r, (The above $0$ axis $e_{+}^{\infty}$ and the lower $0$ axis $e_{-}^{\infty}$)}
\label{Fig:3}
\end{figure*}

The Fig. \ref{Fig:3} represents the negative energy density at infinity with magnetization ($\sigma)$ for different values of CoS parameter with fixed values of ($g=0.1,\, \& \,\xi=0,\pi/18$).  The effect of the CoS parameter on the negative energy density at infinity is small but significant. We can see that $e_{+}^{\infty}$ increases and   $e_{-}^{\infty}$ decreases with the plasma magnetization ($\sigma$). 

\begin{figure*}[ht]
\begin{tabular}{c c c c}
\includegraphics[width=0.45\linewidth]{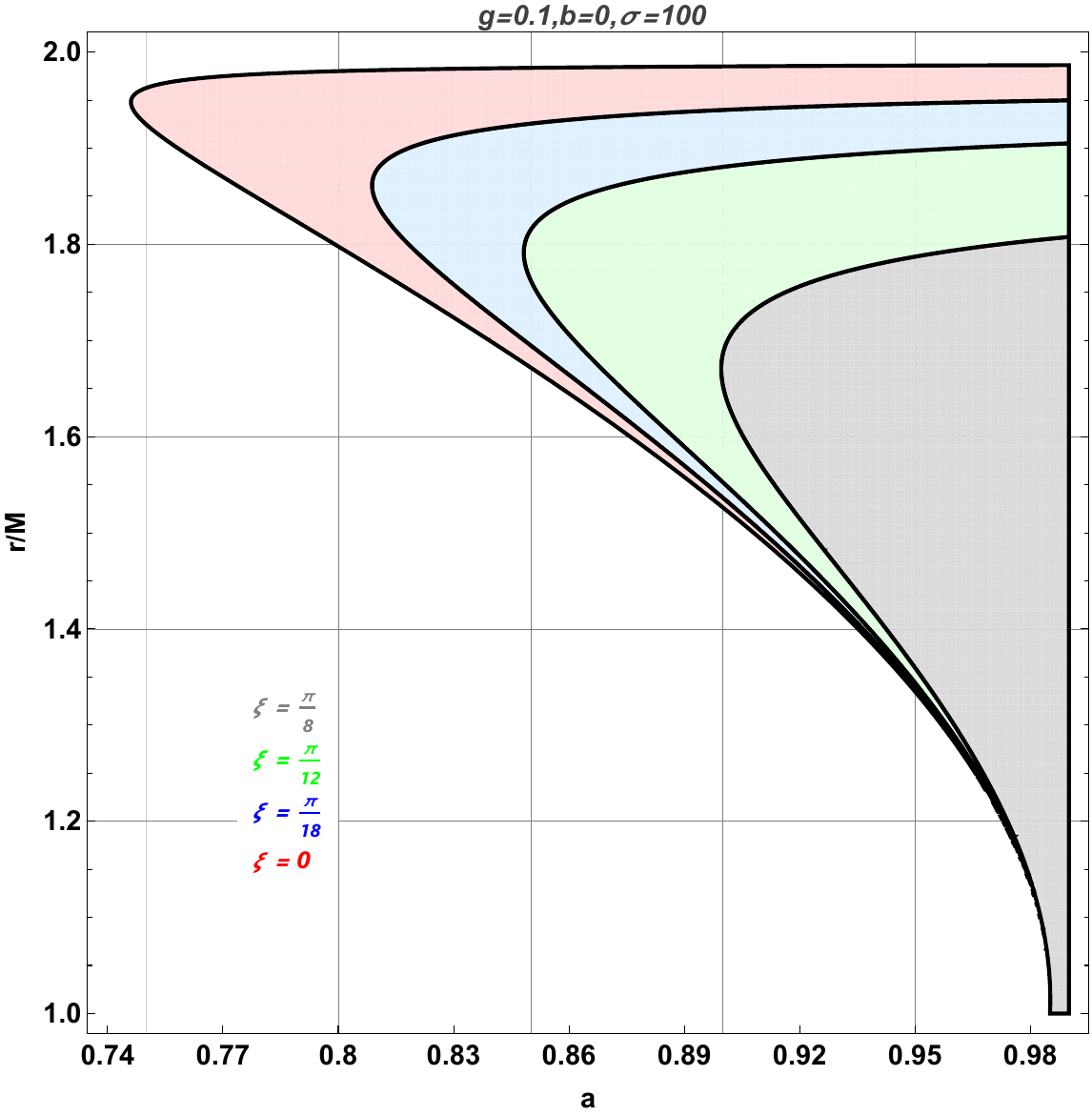}
\hspace{0.5 cm}
\includegraphics[width=0.45\linewidth]{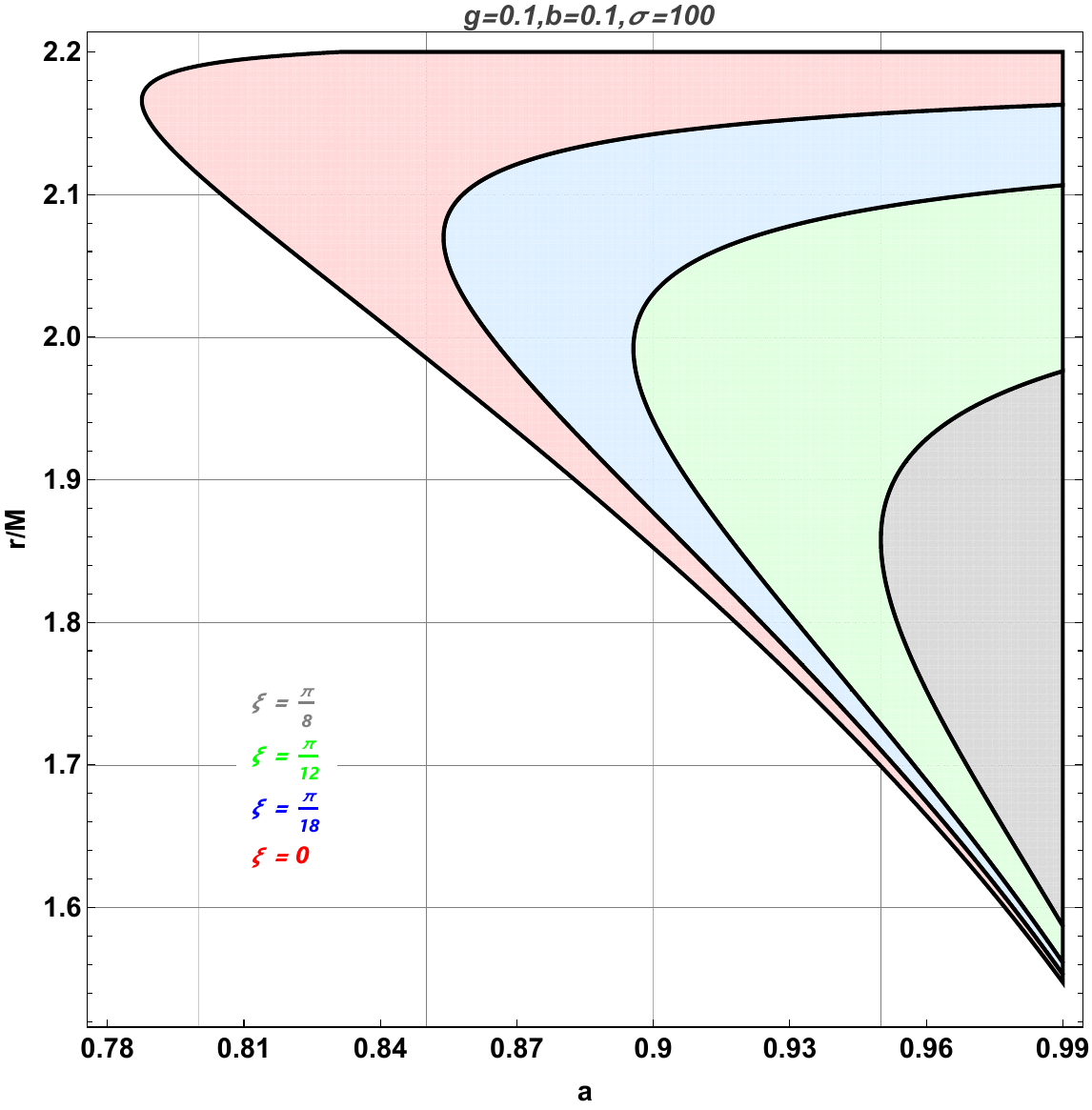}
\end{tabular}
\caption{Plot of phase space ($r/M$ \text{vs} $a$) for different values of $\xi$ with fixed value of $\sigma\ \& \ g=0.1$.} 
\label{Fig:4}
\end{figure*}

Let us examine the parameter space of the radial location ($r/M$) and the black hole spin ($a$), allowing energy extraction from the Black hole via magnetic reconnection. The results are depicted in Fig. \ref{Fig:4} for different values of plasma magnetization and with fixed values of orientation angle ($\xi=\pi/18$) corresponding to the fixed value of the CoS parameter ($b=0,\ b=0.1,\ b=0.2 \& \ b=0.3$). The peaks are shifted to higher values of ($r/M$) for increasing the values of the CoS parameter ($b=0 \to 0.3$), and the allowed region of the phase space of the negative energy region ($e_{-}^{\infty}<0$) decreases with increasing the value of the CoS parameter.

\begin{figure*}[ht]
\begin{tabular}{c c}
\includegraphics[width=.45\linewidth]{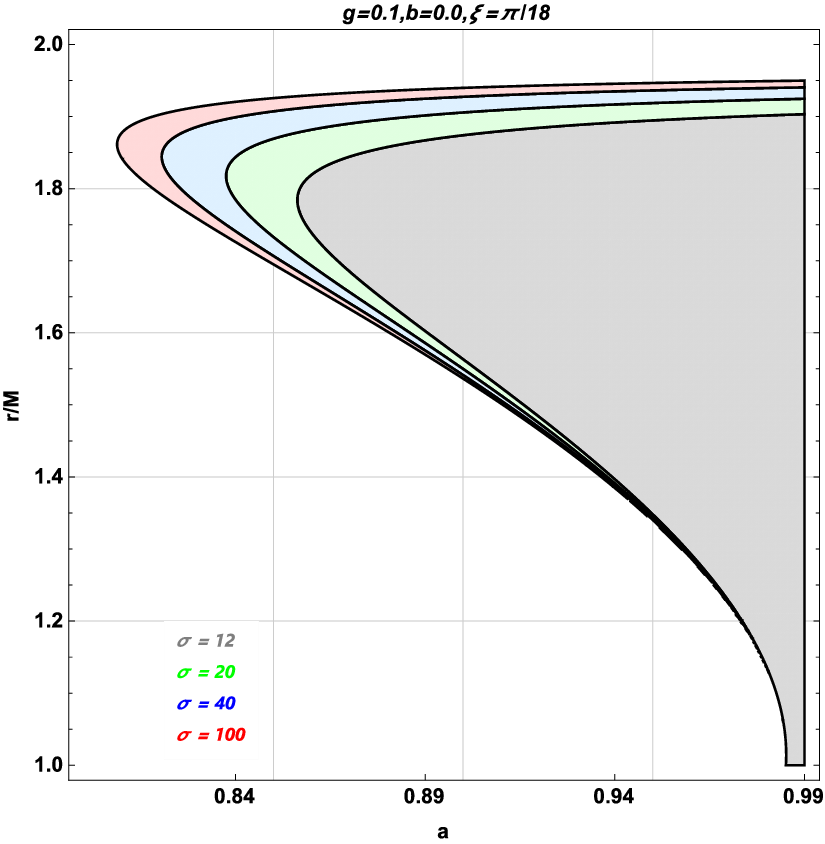}
\hspace{0.5 cm}
\includegraphics[width=.45\linewidth]{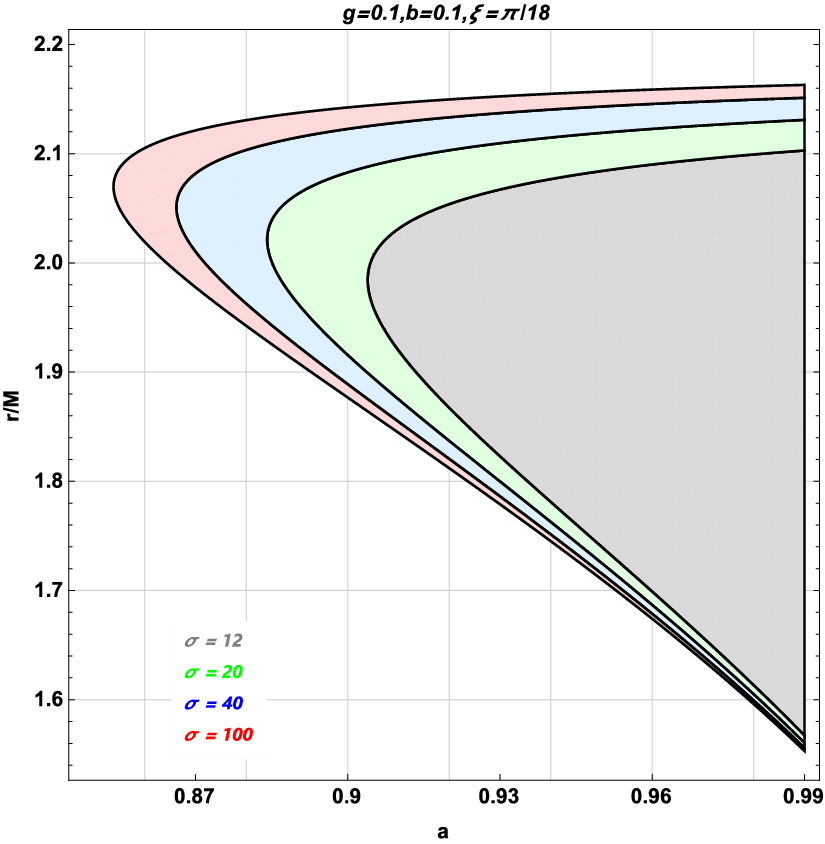}
\end{tabular}
\caption{Plot of phase space ($r/M$ \text{vs} $a$) for different values of $\sigma$ corresponding to fixed value of  $g=0.1$ and ($b = 0.0,\, b=0.1$) }
\label{Fig:5}
\end{figure*}

\vspace{9mm}
%%%%%%%%%%%%%%%%%%%%%%%%%%%%%%%%%%%
\section{\label{sec:level4} ENERGY EXTRACTION AND EFFICIENCY}
 In this section, we study the rate of energy extraction and the efficiency of the Kerr-Bardeen-Letelier black hole using the Comisso-Asenjo mechanism. The rate of energy extraction or power is given by \cite{Comisso21}
\begin{eqnarray}
    P_{Energy}=-{\epsilon}_{-}^{\infty} w {A_{0}} {U}_{in},
\end{eqnarray}
where ${A_{0}}\approx r_{e}^{2}-r_{ph}^{2}$ is the cross-sectional area for the inflowing plasma of a rotating black hole, $r_e$ is the outer ergo-sphere and $r_{ph}$ is the radius of the photon sphere. The  ${U}_{in}$  is the reconnection inflow four velocity $O(10^{-1})$ referring to the collisionless regime \cite{lin17,Com16,Com19} and $O(10^{-7})$ referring to the collisional regime \cite{Uzd10,Com2016,ks16}.

\begin{figure*}[ht]
\begin{tabular}{c c}
\includegraphics[width=.45\textwidth]{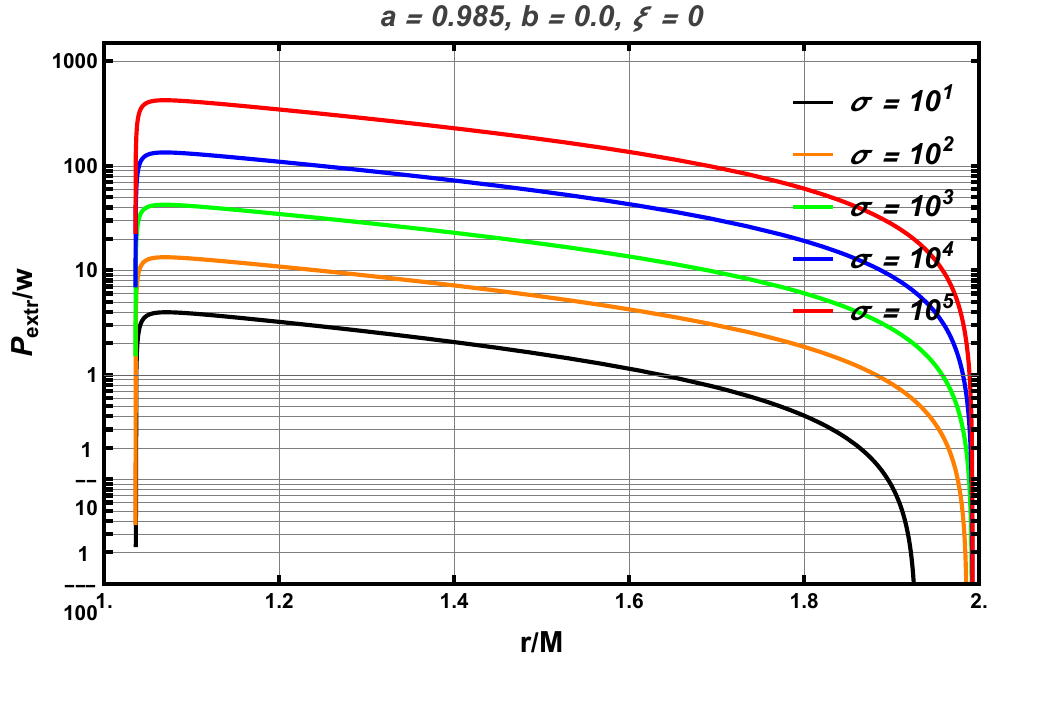}
\includegraphics[width=.45\textwidth]{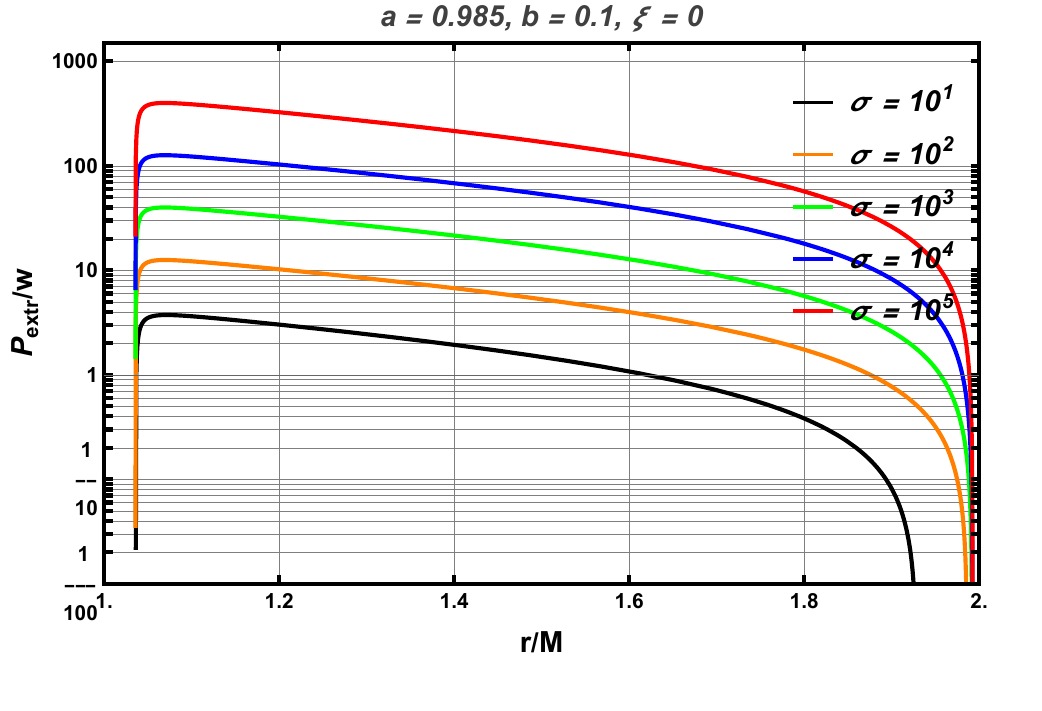}
\end{tabular}
\caption{Plots of power extraction rate per enthalpy vs  $r/M$ for different plasma magnetization ($\sigma$) at fixed magnetic monopole charge $g=0.1$.}
\label{Fig:6}
\end{figure*}

The plot of energy extraction rate per unit enthalpy $P_{extr}/{w}$ with different values of plasma magnetization parameter ($\sigma$) is depicted in Fig. \ref{Fig:6}. The power extracted from the black hole solution (\ref{bhs}) increases monotonically with increasing values of $\sigma$.   For small values of $r/m$, $P_{extr}/{w}$ reaches a maximum, then drops off.  The peak of $P_{extr}/{w}$ decreases with increasing values of the CoS parameter (see Fig.  \ref{Fig:6}). We notice that the rate of energy extraction increases with magnetization of plasma. Thus the energy extraction rate is significantly affected by the spin angular momentum of the black hole, and for a higher spin, the $r/M$ point (corresponding to maxima) is shifted towards a smaller value of $r/M$. 

The efficiency of the Comisso-Asenjo mechanism is given by the following expression \cite{Comisso21}
\begin{equation}
    \eta = \frac{\epsilon^{\infty}_{+}}{\epsilon^{\infty}_{+}+\epsilon^{\infty}_{-}}.
    \label{eff}
\end{equation}
The maximum efficiency is obtained at $a\to 1,\, r/M \to 1,\,\xi \to 0$ and $\sigma >>1$. In this limit, the efficiency becomes
\begin{equation}
   \eta\approx \frac{\sqrt{3\sigma}}{\sqrt{3\sigma}-\sqrt{\frac{\sigma}{3}}} =\frac{3}{2}.
\end{equation}
We can see (from Fig.7) that the efficiency decreases with CoS parameter but increases with the spin angular momentum of the black hole.  The energy extraction from the Kerr-Bardeen-Letelier balck hole is dominated in the high spin range ($0.95 \to 0.98$). 

\begin{figure*}[ht]
\begin{tabular}{c c}
 \includegraphics[width=.45\textwidth]{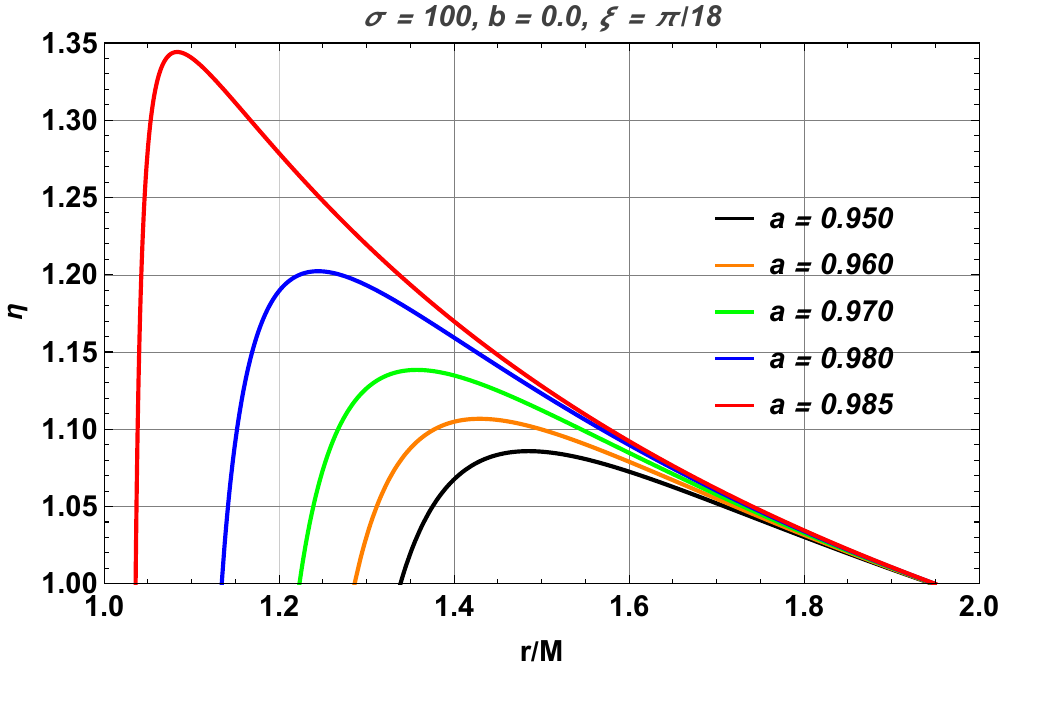}
 \includegraphics[width=.45\textwidth]{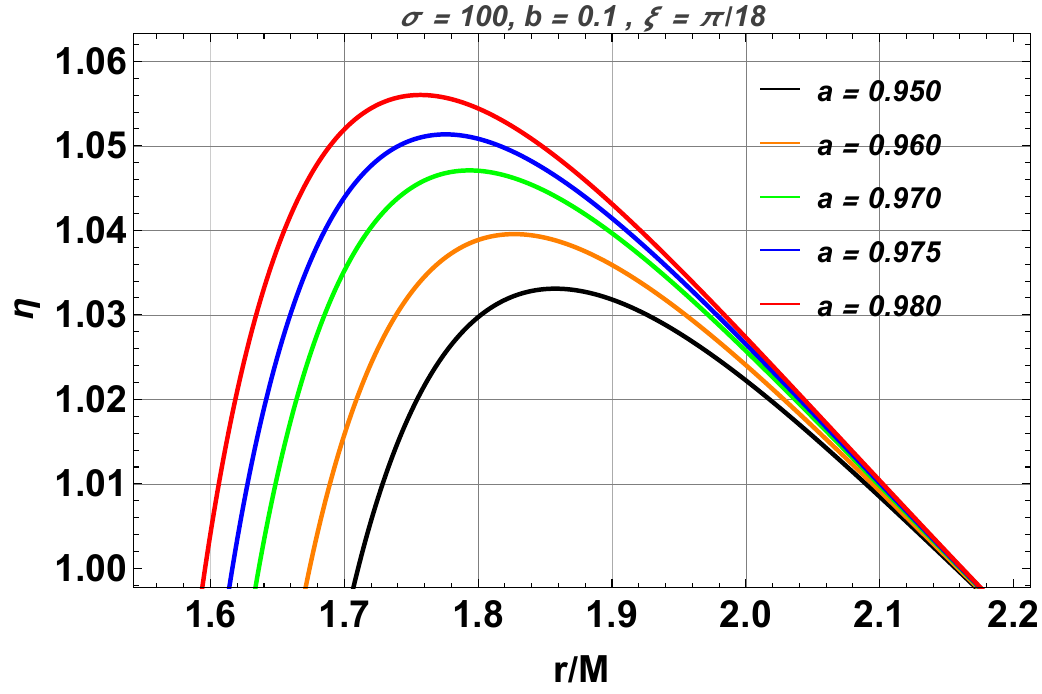}
  	\end{tabular}
   \caption{Plot of efficiency of magnetic reconnection vs  $r/M$ for different values of angular momentum of black hole at fixed magnetic monopole charge $g=0.1$.}
\label{Fig:7}
\end{figure*}
 
Now, let us compare the Comisso-Asenjo mechanism with the Blandford-Znajek process of energy extraction. The Blandford-Znajek process elegantly explains the electromagnetic extraction of energy from a rotating black hole because the magnetic field lines thread the ergoregion. The power extracted in Blandford-Znazek process  can be written as,  
\begin{equation}
 P_{BZ}=\kappa \Phi{^2}\Omega^2_{h}\left(\left[1+\lambda_{1}\Omega^2_{h}+\lambda_{2}\Omega^4_{h}\right]\right).  
\end{equation}
$\kappa$  is related to the magnetic field geometry near the ergo-region of black hole, $\Phi$ is the magnetic flux that traversing the hemisphere of the black hole horizon, $\Omega_{h}$ angular frequency of horizon and $\lambda_{1} =1.38$ and $\lambda_{2}=9.2$ are constants \cite{Comisso21}. For the steady state, the Blandford-Znazek process magnetic flux is given as $\Phi_{BZ}\approx ~B_{0} r^2_{h}~sin{\xi}$, with $B^2_{0}=\sigma w$. We can take the ration of the efficiency of the  Comisso-Asenjo mechanism with that of the Blandford-Znazek process, and is given by;
\begin{equation}
    \frac{P_{CA}}{P_{BZ}}\sim\frac{-\epsilon^{\infty}_{-}A_{in}U_{in}}{\kappa \Omega^{2}_{h}r^2_{h}\sigma sin^2 \xi\left[1+\lambda_{1}\Omega^2_{h}+\lambda_{2}\Omega^4_{h}\right]}.
\end{equation}

\begin{figure*}[ht]
\begin{tabular}{c c}
\includegraphics[width=.45\textwidth]{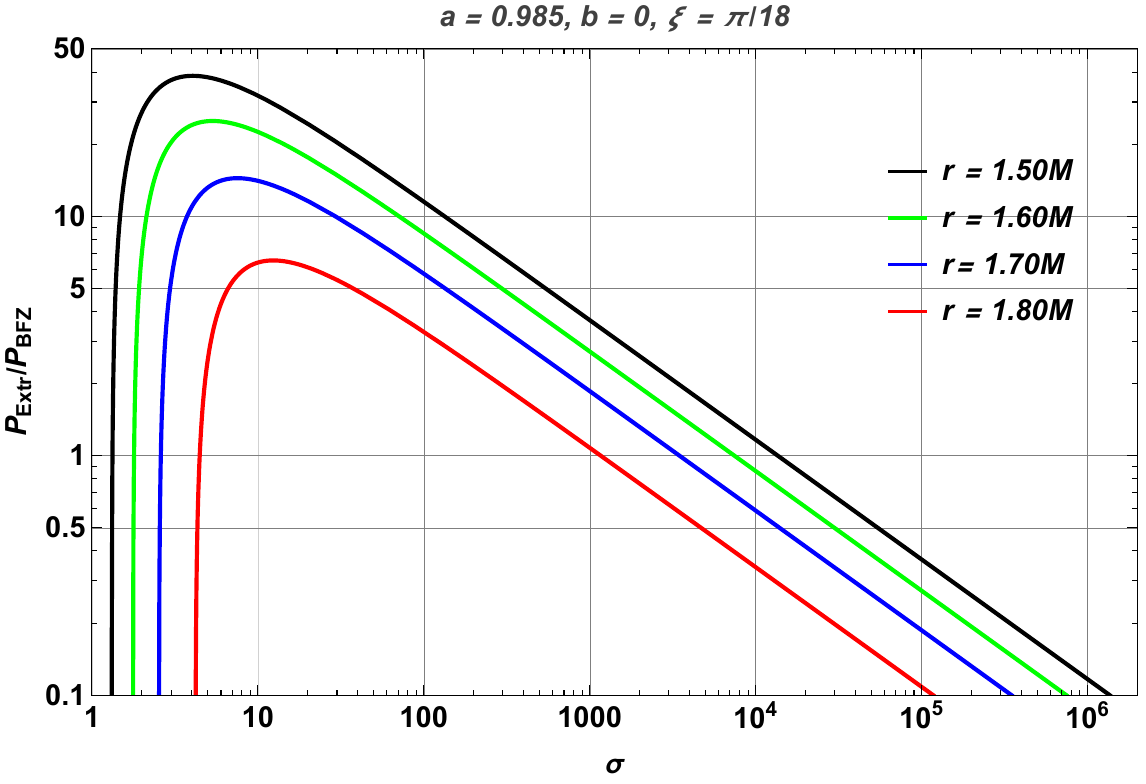}
\includegraphics[width=.45\textwidth]{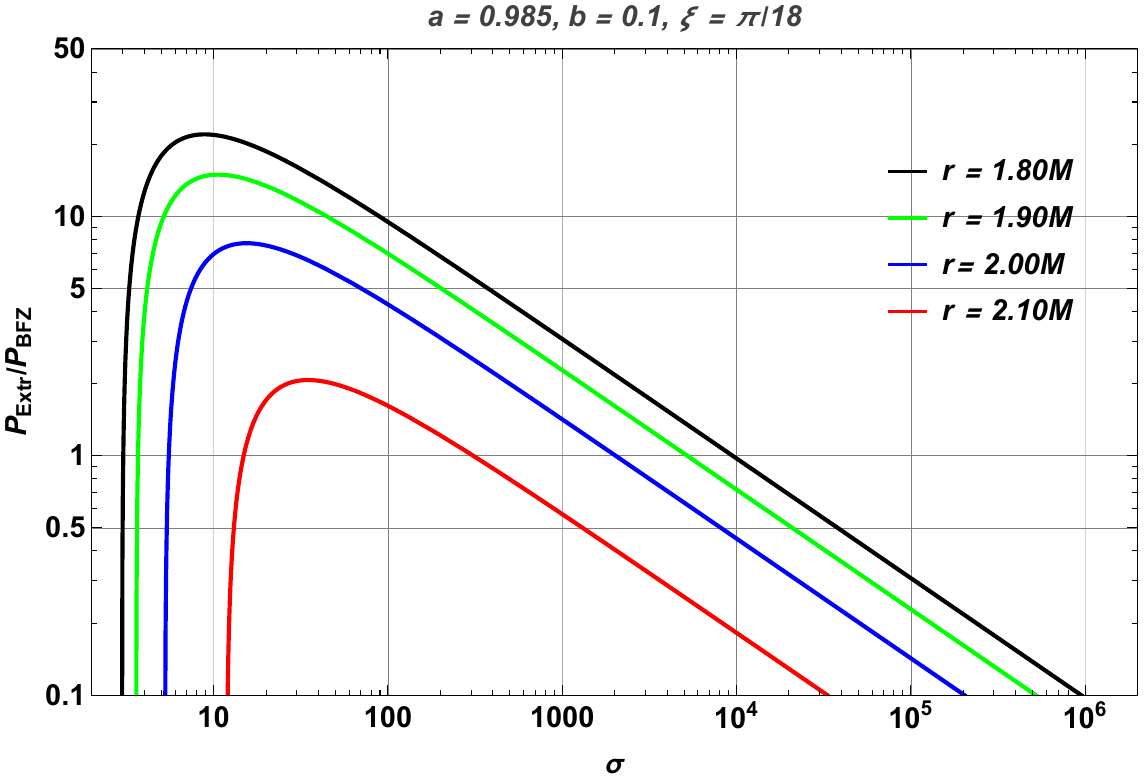}
\end{tabular}
\caption{The plot of power extraction relative to the Blandford-Znajek mechanism vs plasma magnetization ($\sigma$) at fixed magnetic charge $g=0.1$.}
\label{Fig:8}
\end{figure*}

 The graph of the power extraction between Comisso-Asenjo  and Blandford-Znajek mechanism vs plasma magnetization ($\sigma$) is plotted in Fig. \ref{Fig:8}. It is clear that energy extraction is more efficient in comparison with the original Blandford-Znajek process in magnetic reconnection. Hence, for rapid magnetic reconnection ${P_{CA}}/{P_{BZ}}$ must be larger than unity.  

\section{\label{sec:level5}SUMMARY AND CONCLUSIONS}
In this paper, we have studied the energy extraction from a Kerr-Bardeen-Letelier black hole via the Comisso-Asenjo mechanism. The plasma energy at infinity is associated with the accelerated and de-accelerated part of reconnection process of the energy extraction from the ergo-region of the black hole. The energy extraction via magnetic reconnection is possible for higher value of black hole spin ($a\sim 1$) and magnetization ($\sigma>1/3$). We analyzed the effect of plasma magnetization ($\sigma$), orientation angle ($\xi$) and CoS parameter ($b$) on plasma energy at infinity and the efficiency of energy extraction. The efficiency of the energy extraction decreases with CoS parameter $(b)$. Furthermore, we also compared the power efficiency between Comisso-Asenjo and Blandford-Znajek mechanism vs plasma magnetization. In this comparison, we observed  that the Comisso-Asenjo  is more powerful than Blandford-Znajek mechanism for energy extraction. 

\begin{acknowledgements}
\noindent
 BKV acknowledges financial support from the UGC fellowship. DVS thanks the Council of Science and Technology (CST) Uttar Pradesh for Grant No. CST/D-828.
\end{acknowledgements}

\end{document}